\documentclass{elsart}
\usepackage{amssymb}
\def\tr{\mathop{\mathrm{Tr}}}
\def\re{\mathop{\mathrm{Re}}}

\begin{document}
\begin{frontmatter}
\title{Influence of the finite duration of the measurement on the quantum Zeno
effect}
\author{Julius Ruseckas}
\ead{ruseckas@itpa.lt}
\address{Institute of Theoretical Physics and Astronomy,\\
A. Go\v{s}tauto 12, 2600 Vilnius, Lithuania}

\begin{abstract}
  We analyze the influence of the finite duration of the measurement on the
  quantum Zeno effect, using a simple model of the measurement. It is shown that
  the influence of the finite duration of the measurement is uninportant
  when this duration is small compared to the duration of the free evolution
  between the measurements.
\end{abstract}
\end{frontmatter}

\section{Introduction}
\label{sec:introd}

The quantum Zeno effect is a consequence of the influence of the measurements on
the evolution of a quantum system. In quantum mechanics the short-time behavior
of the non-decay probability of unstable particle is not exponential but
quadratic \cite{Khalfin}. This deviation from the exponential decay has been
observed experimentally \cite{GG,newexp}. In 1977, Mishra and Sudarshan
\cite{Mishra} showed that this behavior when combined with the quantum theory of
measurement, based on the assumption of the collapse of the wave function, led
to a very surprising conclusion: frequent observations slowed down the decay.
They modeled the continuous observation of the system by a succession of the
instantaneous measurements with free evolution of the system between the
measurements.

Cook \cite{Cook} suggested an experiment on the quantum Zeno effect that was
realized by Itano {\it et al.\/} \cite{Itano}. In this experiment a repeatedly
measured two-level system undergoing Rabi oscillations has been used.  The
outcome of this experiment has also been explained without the collapse
hypothesis \cite{Petrosky,FS,Namiki}. Recently, an experiment similar to Ref.\ 
\cite{Itano} has been performed by Balzer {\it et al.\/} \cite{Balzer}. The
quantum Zeno and anti-Zeno effects have been experimentally observed in Ref.\ 
\cite{newexp}

In the analysis of the quantum Zeno effect the finite duration of the
measurement becomes important. In Ref.\ \cite{rus2} a simple model that allows
us to take into account the finite duration and finite accuracy of the
measurement has been developed. However, in Ref.\ \cite{rus2} it has been
analyzed the case when there are no free evolution between the measurements. In
this article we obtain the corrections to the jump probability due to the finite
duration of the measurement with the free evolution between the measurements.

We proceed as follows: In Sec.~\ref{sec:model} we present the model of the
measurement. Sec.~\ref{sec:decay-rate} is devoted to the derivation of the
formula for the probability of the jump into another level during the
measurement of the frequently measured perturbed system. 
In Sec. \ref{sec:example} the evolution of the measured two-level system is
analysed as an example of the application of
our model. Sec.~\ref{sec:concl} summarizes our findings.

\section{Model of the measurements}
\label{sec:model}

We consider a system which consists of two parts. The first part of the system
has the discrete energy spectrum. The Hamiltonian of this part is $\hat{H}_0$.
The other part of the system is represented by Hamiltonian $\hat{H}_1$.
Hamiltonian $\hat{H}_1$ commutes with $\hat{H}_0$. In a particular case the
second part may be absent and $\hat{H}_1$ may be zero. The operator
$\hat{V}(t)$ causes the jumps between the different energy levels of $\hat{H}_0$.
Therefore, the full Hamiltonian of the system equals to
$\hat{H}_S=\hat{H}_0+\hat{H}_1+\hat{V}(t)$. An example of such a system is an
atom with the Hamiltonian $\hat{H}_0$ interacting with the electromagnetic
field, represented by $\hat{H}_1$.

We will measure in which eigenstate of the Hamiltonian $\hat{H}_0$ the
system is. The measurement is performed by coupling the system with the
detector. The full Hamiltonian of the system and the detector equals to
\begin{equation}
\hat{H}=\hat{H}_{\mathrm{S}}+\hat{H}_{\mathrm{D}}+\hat{H}_{\mathrm{I}} ,
\label{eq:first}
\end{equation}
where $\hat{H}_{\mathrm{D}}$ is the Hamiltonian of the detector and
$\hat{H}_{\mathrm{I}}$ represents the interaction between the detector and the
system. We choose the operator $\hat{H}_{\mathrm{I}}$ in the form
\begin{equation}
\hat{H}_{\mathrm{I}}=\lambda\hat{q}\hat{H}_0 ,
\end{equation}
where $\hat{q}$ is the operator acting in the Hilbert space of the detector and
the parameter $\lambda$ describes the strength of the interaction. This
system--detector interaction is considered by von Neumann \cite{vNeum} and in
Refs.\ \cite{rus2,joos,caves,milb,gagen,rus,rus3}. In order to obtain a
sensible measurement, the parameter $\lambda$ must be large. We require a
continuous spectrum of operator $\hat{q}$. For simplicity, we can consider the
quantity $q$ as the coordinate of the detector.

The measurement begins at time moment $t_0$. At the beginning of the interaction
with the detector, the detector is in the pure state $|\Phi\rangle$. The full
density matrix of the system and detector is $\hat{\rho}(t_0)=
\hat{\rho}_{\mathrm{S}}(t_0)\otimes|\Phi\rangle\langle\Phi|$ where
$\hat{\rho}_{\mathrm{S}}(t_0)$ is the density matrix of the system. The duration
of the measurement is $\tau$. After the measurement the density matrix of the
system is $\hat{\rho}_{\mathrm{S}}(\tau+t_0)=\tr_{\mathrm{D}}\{
\hat{U}_{\mathrm{M}}(\tau,t_0)(\hat{\rho}_{\mathrm{S}}(t_0)\otimes
|\Phi\rangle\langle\Phi|)\hat{U}_{\mathrm{M}}^{\dag}(\tau,t_0)\}$
where $\hat{U}_{\mathrm{M}}(t,t_0)$ is the evolution operator of
the system and detector, obeying the equation
\begin{equation}
\mathrm{i}\hbar\frac{\partial}{\partial t}\hat{U}_{\mathrm{M}}(t,t_0)=
\hat{H}(t+t_0)\hat{U}_{\mathrm{M}}(t,t_0)
\end{equation}
with the initial condition $\hat{U}_{\mathrm{M}}(0,t_0)=1$. Further, for
simplicity we will neglect the Hamiltonian of the detector (as in Ref.\ 
\cite{rus2}). Then the evolution operator $\hat{U}_{\mathrm{M}}$ obeys the
equation
\begin{equation}
\mathrm{i}\hbar\frac{\partial}{\partial t}\hat{U}_{\mathrm{M}}(t,t_0)=
\left((1+\lambda\hat{q})\hat{H}_0+\hat{H}_1+\hat{V}(t+t_0)\right)
\hat{U}_{\mathrm{M}}(t,t_0) .
\label{eq:evol1}
\end{equation}

After the measurement the system is left for the measurement-free evolution for
time $T-\tau$. The density matrix becomes $\hat{\rho}_{\mathrm{S}}(T+t_0)=
\hat{U}_{\mathrm{F}}(T-\tau,\tau+t_0)\hat{\rho}_{\mathrm{S}}(\tau+t_0)
\hat{U}_{\mathrm{F}}^{\dag}(T-\tau,\tau+t_0)\}$, where
$\hat{U}_{\mathrm{F}}(t,t_0)$ is the evolution operator of the system only,
obeying the equation
\begin{equation}
\mathrm{i}\hbar\frac{\partial}{\partial t}\hat{U}_{\mathrm{F}}(t,t_0)=
\hat{H}_{\mathrm{S}}(t+t_0)\hat{U}_{\mathrm{F}}(t,t_0)
\end{equation}
with the initial condition $\hat{U}_{\mathrm{F}}(0,t_0)=1$.

The measurements of the duration $\tau$ with a subsequent free evolution for the
time $T-\tau$ are repeated many times with the measurement period $T$. Such a
process was considered by the Mishra and Sudarshan \cite{Mishra} and realized in
the experiments \cite{Itano}.

\section{Jump probability}
\label{sec:decay-rate}

We will calculate the probability of the jump from the initial to the final
state during the measurement and subsequent measurement-free evolution. The
jumps are induced by the operator $\hat{V}(t)$ that represents the perturbation
of the unperturbed Hamiltonian $\hat{H}_0+\hat{H}_1$. We will take into account
the influence of the operator $\hat{V}$ by the perturbation method, assuming
that the durations of the measurement $\tau$ and of the free evolution $T-\tau$
are small.

The operator $\hat{V}(t)$ in the interaction picture during the measurement is
\begin{equation}
\tilde{V}_{\mathrm{M}}(t,t_0)=\hat{U}^{(0)}_{\mathrm{M}}(t)\hat{V}(t+t_0)
\hat{U}^{(0)}_{\mathrm{M}}(t) ,
\end{equation}
where $\hat{U}^{(0)}_{\mathrm{M}}(t)$ is the evolution operator of the system
and the detector (\ref{eq:first}) without the perturbation $\hat{V}$
\begin{equation}
\hat{U}^{(0)}_{\mathrm{M}}(t)=\exp\left(-\frac{\mathrm{i}}{\hbar}(\hat{H}_0
+\hat{H}_1+\hat{H}_{\mathrm{I}})t\right) .
\label{eq:1}
\end{equation}
The evolution operator $\hat{U}_{\mathrm{M}}(\tau,t_0)$ in the second order
approximation equals to
\begin{eqnarray}
\hat{U}_{\mathrm{M}}(\tau,t_0)&\approx&\hat{U}^{(0)}_{\mathrm{M}}(\tau)\left(1+
\frac{1}{\mathrm{i}\hbar}\int_0^\tau\d t\tilde{V}_{\mathrm{M}}(t,t_0)
\right.\nonumber\\
&&-\left.\frac{1}{\hbar^2}\int_0^\tau\d t_1\int_0^t\d t_2\tilde{V}_{\mathrm{M}}
(t_1,t_0)\tilde{V}_{\mathrm{M}}(t_2,t_0)\right).
\end{eqnarray}

The operator $\hat{V}(t)$ in the interaction picture during the free evolution
is
\begin{equation}
\tilde{V}_{\mathrm{F}}(t,t_0)=\hat{U}^{(0)}_{\mathrm{F}}(t)\hat{V}(t+t_0)
\hat{U}^{(0)}_{\mathrm{F}}(t) ,
\end{equation}
where $\hat{U}^{(0)}_{\mathrm{F}}(t)$ is the evolution operator of the system
without the perturbation $\hat{V}$, i.e.,
\begin{equation}
\hat{U}^{(0)}_{\mathrm{F}}(t)=\exp\left(-\frac{\mathrm{i}}{\hbar}(\hat{H}_0+
\hat{H}_1)t\right) .
\label{eq:2}
\end{equation}
The evolution operator $\hat{U}_{\mathrm{F}}(t,t_0)$ in the second order
approximation equals to
\begin{eqnarray}
\hat{U}_{\mathrm{F}}(t,t_0)&\approx&\hat{U}^{(0)}_{\mathrm{F}}(t)\left(1+
\frac{1}{\mathrm{i}\hbar}\int_0^t\d t_1\tilde{V}_{\mathrm{F}}(t_1,t_0)
\right.\nonumber\\
&&-\left.\frac{1}{\hbar^2}\int_0^t\d t_1\int_0^t\d t_2\tilde{V}_{\mathrm{F}}(t_1,t_0)
\tilde{V}_{\mathrm{F}}(t_2,t_0)\right).
\end{eqnarray}

We can choose the basis $\left|n\alpha\right\rangle$ common for the operators
$\hat{H}_0$ and $\hat{H}_1$,
\begin{eqnarray}
\hat{H}_0\left|n\alpha\right\rangle&=&E_n\left|n\alpha\right\rangle , \\
\hat{H}_1\left|n\alpha\right\rangle&=&E_1(n,\alpha)\left|n\alpha\right\rangle ,
\end{eqnarray}
where $n$ numbers the eigenvalues of the Hamiltonian $\hat{H}_0$ and $\alpha$
represents the remaining quantum numbers.

The probability of the jump from the level $|i\alpha\rangle$ to the level
$|f\alpha_1\rangle$ is 
\begin{eqnarray}
W(i\alpha\rightarrow f\alpha_1)&=&\tr\nolimits_{\mathrm{D}}\{\langle f\alpha_1|
\hat{U}_{\mathrm{F}}(T-\tau)\hat{U}_{\mathrm{M}}(\tau)
(|i\alpha\rangle\langle i\alpha|\otimes|\Phi\rangle\langle\Phi|)
\nonumber\\
&&\times\hat{U}_{\mathrm{F}}^{\dag}(T-\tau)\hat{U}_{\mathrm{M}}^{\dag}(\tau)
|f\alpha_1\rangle\} .
\label{eq:3}
\end{eqnarray}

In the second-order approximation we obtain the expression for the jump
probability $W(i\alpha\rightarrow f\alpha_1)$. The jump probability consists
from three parts.
\begin{equation}
W(i\alpha\rightarrow f\alpha_1)=W_{\mathrm{F}}(i\alpha\rightarrow f\alpha_1)+
W_{\mathrm{M}}(i\alpha\rightarrow f\alpha_1)+
W_{\mathrm{Int}}(i\alpha\rightarrow f\alpha_1) ,
\label{eq:6}
\end{equation}
where $W_{\mathrm{F}}$ is the probability of the jump during the free evolution,
$W_{\mathrm{M}}$ is the probability of the jump during the measurement and
$W_{\mathrm{Int}}$ is an interference term.  The expressions for these
probabilities are (see Refs.\ \cite{rus2,rus} for the analogy of the derivation)
\begin{eqnarray}
W_{\mathrm{F}}(i\alpha\rightarrow f\alpha_1)&=&\frac{1}{\hbar^2}
\int_0^{T-\tau}\d t_1 \int_0^{T-\tau}\d t_2V(t_1+t_0+\tau)_{f\alpha_1,i\alpha}
V(t_2+t_0+\tau)_{i\alpha,f\alpha_1}\nonumber\\
&&\times\exp(\mathrm{i}\omega_{f\alpha_1,i\alpha}(t_1-t_2)),
\label{eq:free} \\
W_{\mathrm{M}}(i\alpha\rightarrow f\alpha_1)&=&\frac{1}{\hbar^2}
\int_0^{\tau}\d t_1 \int_0^{\tau}\d t_2V(t_1+t_0)_{f\alpha_1,i\alpha}
V(t_2+t_0)_{i\alpha,f\alpha_1}\nonumber\\
&&\times\exp(\mathrm{i}\omega_{f\alpha_1,i\alpha}(t_1-t_2))
F(\lambda\omega_{fi}(t_1-t_2)) ,
\label{eq:measur} \\
W_{\mathrm{Int}}(i\alpha\rightarrow f\alpha_1)&=&\frac{2}{\hbar^2}\re
\int_0^{\tau}\d t_1\int_{\tau}^{T}\d t_2V(t_1+t_0)_{f\alpha_1,i\alpha}
V(t_2+t_0)_{i\alpha,f\alpha_1}\nonumber\\
&&\times\exp(\mathrm{i}\omega_{f\alpha_1,i\alpha}(t_1-t_2))F(\lambda\omega_{if}
(\tau-t_1)) ,
\label{eq:interf}
\end{eqnarray}
where
\begin{eqnarray}
\omega_{fi}&=&\frac{1}{\hbar}(E_f-E_i), \\
\omega_{f\alpha_1,i\alpha}&=&\omega_{fi}+\frac{E_1(f,\alpha_1)-E_1(
i,\alpha)}{\hbar},\\
F(x)&=&\langle\Phi|\exp(\mathrm{i}x\hat{q})|\Phi\rangle .
\end{eqnarray}

The probability to remain for the system in the initial state $|i\alpha\rangle$
is
\begin{equation}
  \label{eq:5}
  W(i\alpha)=1-\sum_{f,\alpha_1}W(i\alpha\rightarrow f\alpha_1) .
\end{equation}
After $N$ measurements the probability for the system to survive in the
initial state is equal to $W(i\alpha)^N\approx\exp(-RNT)$,
where $R$ is the measurement-modified decay rate
\begin{equation}
  \label{eq:8}
  R=\sum_{f,\alpha_1}\frac{1}{T}W(i\alpha\rightarrow f\alpha_1)
\end{equation}

\section{Example}
\label{sec:example}

As an example we will consider the evolution of the measured two-level system.
The system is forced by the periodic of the frequency $\omega_L$ perturbation
$V(t)$ which induces the jumps from one state to another. Such a system was used
in the experiment by Itano {\it et al\/} \cite{Itano}. The Hamiltonian of this
system is
\begin{equation}
\hat{H}=\hat{H}_0+\hat{V}(t)
\label{eq:ham1}
\end{equation}
where
\begin{eqnarray}
\hat{H}_0&=&\frac{\hbar\omega}{2}\hat{\sigma}_3, \label{eq:ham2}\\
\hat{V}(t)&=&(v\hat{\sigma}_{+}+v^{*}\hat{\sigma}_{-})
\cos(\omega_{\mathrm{L}} t).
\label{eq:ham3}
\end{eqnarray}
Here $\sigma_1,\sigma_2,\sigma_3$ are Pauli matrices and $\sigma_{\pm}
=\frac{1}{2}(\sigma_1\pm i\sigma_2)$. The Hamiltonian $\hat{H}_0$ has two
eigenfunctions $|0\rangle$ and $|1\rangle$ with the eigenvalues
$-\hbar\frac{\omega}{2}$ and $\hbar\frac{\omega}{2}$ respectively.

Using Eqs.\ (\ref{eq:free}), (\ref{eq:measur}) and (\ref{eq:interf}) for the
jump from the state $|0\rangle$ to the state $|1\rangle$ we obtain
\begin{eqnarray}
W_{\mathrm{F}}(0\rightarrow 1)&=&\frac{|v|^2}{\hbar^2}\frac{\sin^2\left(
\frac{\Delta\omega}{2}(T-\tau)\right)}{(\Delta\omega)^2} ,\\
W_{\mathrm{M}}(0\rightarrow 1)&=&\frac{\tau}{2}\frac{|v|^2}{\hbar^2}\re
\int_0^{\tau}F(\lambda\omega t)\exp(\mathrm{i}\Delta\omega t)
\left(1-\frac{t}{\tau}\right)\d t , \label{eq:WM}\\
W_{\mathrm{Int}}(0\rightarrow 1)&=&\frac{|v|^2}{2\hbar^2}\re\int_0^{\tau}\d t_1
\int_{\tau}^{T}\d t_2\exp(\mathrm{i}\Delta\omega(t_1-t_2))
F(\lambda\omega(t_1-\tau)) ,
\label{eq:WInt}
\end{eqnarray}
where $\Delta\omega=\omega-\omega_{\mathrm{L}}$ is the detuning . Equation
(\ref{eq:WM}) has been obtained in Ref.\ \cite{rus2}.

When $\lambda$ is large, the function $F$ varies rapidly and we can approximate
expressions (\ref{eq:WM}) and (\ref{eq:WInt}) as
\begin{eqnarray}
W_{\mathrm{M}}(0\rightarrow 1)&=&\frac{\tau}{2\Lambda\omega}
\frac{|v|^2}{\hbar^2}\\ 
W_{\mathrm{Int}}(0\rightarrow 1)&=&\frac{|v|^2}{\hbar^2}
\frac{1}{2\Lambda\omega\Delta\omega}\sin(\Delta\omega(T-\tau))
\end{eqnarray}
where $\Lambda=\lambda/C$, $C$ is the width of the function $F$, defined by the
equation (see Ref.\ \cite{rus2})
\begin{equation}
C=\frac{1}{2}\int_{-\infty}^{\infty}F(x)\d x
\end{equation}
If $T\gg\tau$ and $\Delta\omega T\ll 1$ then we obtain
\begin{equation}
W(0\rightarrow 1)=\frac{|v|^2}{\hbar^2}\frac{T^2}{4}+
\frac{|v|^2}{\hbar^2}\frac{T}{2}\left(\frac{1}{\Lambda\omega}-\tau\right) .
\label{eq:result}
\end{equation}

From Eq.\ (\ref{eq:result}) we see that the jump probability for the non-ideal
measurement consists of two terms. The first term equals to the jump probability
when the measurement is instantaneous, the second term represents the correction
due to the finite duration of the measurement. In Ref.\ \cite{rus2} it has been
shown that the duration of the measurement can be estimated as
\begin{equation}
  \label{eq:4}
  \tau\gtrsim\frac{1}{\Lambda\omega} .
\end{equation}
From Eq.\ (\ref{eq:result}) we see that the correction term is small, since the
duration of the measurement $\tau$ is almost compensated by the term
$1/\Lambda\omega$.

\section{Conclusion}
\label{sec:concl}

The quantum Zeno effect is often analysed using the succession of the
instantaneous measurements with free evolution of the measured system between
the measurements. We analyze here the measurements with finite duration,
instead. We apply the model of the measurement, developed in Ref.\ \cite{rus2}.
The equations for the jump probability (\ref{eq:6})-(\ref{eq:interf}) are
obtained.  Applying the equations to the measured two-level system we obtain a
simple expression for the probability of the jump from one level to the other
(\ref{eq:result}). The influence of the finite duration of the measurement is
expressed as the small correction.

\begin{ack}
  I wish to thank Prof. B. Kaulakys for his suggestion of the problem,
  stimulating discussions and critical remarks.
\end{ack}

\end{document}